\documentclass[]{aa}
\usepackage{graphics}

\begin{document}

\thesaurus{04         
           (04.19.1); 
           08         
           (08.12.2;  
            08.16.5); 
           10         
            (10.15.2)}

\title{Brown Dwarfs in the Pleiades Cluster.}

\subtitle{III. A deep $IZ$ survey\thanks{Based on observations 
          made with the Isaac Newton Teles\-cope (INT) and the Jacobus 
          Kaptein Telescope (JKT) operated on the island of La Palma 
          by the Isaac Newton Group at the Observatorio del Roque de los 
          Muchachos of the Instituto de Astrof\'\i sica de Canarias; 
          and on observations made with the Nordic Optical Telescope 
          (NOT) at the same observatory.}
         }

\author{M. R. Zapatero Osorio\inst{1}, R. Rebolo\inst{1}, 
       E. L. Mart\'\i n\inst{1,}\thanks{\emph{Present address:} 
                 Astronomy Department, University of California, Berkeley, 
                 CA 94720, USA}, 
       S. T. Hodgkin\inst{2}, M. R. Cossburn\inst{2}, 
       A. Magazz\`u\inst{3,}\thanks{\emph{Also at:} Osservatorio Astrofisico 
                 di Catania, Citt\`a Universitaria, Viale Andrea Doria 6, 
                 I-95125 Catania. Italy}, 
       I. A. Steele\inst{4},
       \and
       R. F. Jameson\inst{2}
       }

\offprints{M.R. Zapatero Osorio}
\mail {mosorio@ll.iac.es}

\institute{Instituto de Astrof\'\i sica de Canarias. C/. V\'\i a L\'actea 
           s/n, E-38200 La Laguna, Tenerife. Spain
      \and Astronomy Group, Department of Physics and Astronomy, 
           University of Leicester, Leicester LE1 7RH. UK
      \and Centro Galileo Galilei, Apartado 565, 
           E-38700 Santa Cruz de La Palma, Islas Canarias. Spain
      \and Astrophysics Research Institute, Liverpool John Moores University, 
          Liverpool L3 3AF. UK
          }

\date{Received ; accepted }

\authorrunning{Zapatero Osorio et al.}

\titlerunning{Brown Dwarfs in the Pleiades Open Cluster. III. A deep $IZ$ 
survey}

\maketitle

\begin{abstract}

We present the results of a deep CCD-based $IZ$ photometric survey of a 
$\sim$\,1\,deg$^2$ area in the central region of the Pleiades Galactic 
open cluster. The magnitude coverage of our survey (from $I\,\sim\,17.5$ 
down to $22$) allows us to detect substellar candidates with masses 
between 0.075 and 0.03\,$M_{\odot}$. Details of the photometric reduction 
and selection criteria are given. Finder charts prepared from the $I$-band 
images are provided.

\keywords{Stars: low-mass, brown dwarfs --
          Stars: pre-main sequence --
          Galaxy: open clusters and associations: Pleiades --
          Astronomical data bases: surveys
}
\end{abstract}


\section{Introduction}

The Pleiades star cluster is an ideal hunting ground for substellar 
objects mainly due to its richness of members, young age, proximity 
and scarce interstellar absorption. Taking advantage of these properties, 
several photometric searches aimed at finding brown dwarfs (BDs) have been 
performed during the last decade (see Hambly \cite{hambly98} for a review). 
The recent spectroscopic confirmations of Pleiades objects at the 
stellar-substellar boundary and genuine substellar members (Basri et al. 
\cite{basri96}; Rebolo et al. \cite{rebolo96}; Mart\'\i n et al. 
\cite{martin98}; Stauffer et al. \cite{stauffer98}) previously discovered 
as a result of optical photometric surveys in small areas suggest that a 
numerous population of very low-mass objects may be found in this cluster. 
This encourages future surveys to discover BDs cooler and less massive 
than those (0.075--0.05\,$M_{\odot}$) previously detected by these surveys. 
The Pleiades offers a unique opportunity to establish the observational 
properties of these rather elusive objects and to characterize the initial 
mass function in the substellar mass regime.

\subsection{The survey}

As part of an on-going search for BDs in the Pleiades, we have conducted 
a deep CCD-based $IZ$ survey using the 2.5\,m Isaac Newton Telescope (INT) 
located on the Observatorio del Roque de los Muchachos (ORM, island of La 
Palma). The area covered was 1.05\,deg$^2$ within the central region of 
the cluster (a small fraction of the total area was also observed using 
the $R$ filter). More than 40 faint ($I\ge 17.5$), very red ($I-Z \ge 0.5$) 
objects have been detected down to $I\,\sim\,22$. Their location in the 
colour-magnitude diagram suggests cluster membership. In this paper we 
report on the details of this survey along with the selection criteria. 
We provide $IZ$ magnitudes, coordinates and finder charts for all 
candidates. Preliminary results of this survey were presented in Zapatero 
Osorio et al. (\cite{osorio97a}, \cite{osorio98a}). An extensive discussion 
on the membership of the candidates and derivation of the initial mass 
function will be given in a forthcoming paper (Zapatero Osorio et al. 
\cite{osorio98b}).


\section{Observations and data reduction}

\begin{table}
\caption[]{\label{pholog} Field (100\,arcmin$^2$) center coordinates}
\begin{center}
\small
\begin{tabular}{rrccr}
\hline
        &           &      &       &      \\
\multicolumn{2}{c}{RA (J2000) DEC} &
\multicolumn{1}{c}{Date} &
\multicolumn{1}{c}{Sep.$^{\rm{a}}$} &
\multicolumn{1}{r}{Filters}\\
\multicolumn{1}{c}{($^{\rm{h \ \ m \ \ s}}$)} &
\multicolumn{1}{c}{(\degr \ \ \arcmin \ \ \arcsec)} &
\multicolumn{1}{c}{(1996)} &
\multicolumn{1}{c}{(arcmin)} &
\multicolumn{1}{c}{} \\
        &           &      &       &      \\
\hline
        &           &      &       &      \\
3 43 35 &  24 30 00 & 21 Sep & 53.02 & $ IZ$\\
3 43 43 &  24 45 00 & 12 Feb & 58.69 & $RI$ \hspace*{0.17cm}\\
3 44 15 &  23 40 30 & 21 Sep & 46.15 & $ IZ$\\
3 44 30 &  23 55 00 & 21 Sep & 36.32 & $ IZ$\\
3 44 30 &  24 40 00 & 20 Sep & 47.44 & $ IZ$\\
3 45 00 &  24 37 00 & 20 Sep & 40.54 & $ IZ$\\
3 45 15 &  23 54 35 & 20 Sep & 27.02 & $ IZ$\\
3 45 45 &  24 10 00 & 20 Sep & 17.37 & $ IZ$\\
3 45 45 &  24 46 00 & 20 Sep & 42.55 & $ IZ$\\
3 45 50 &  23 55 00 & 20 Sep & 20.00 & $ IZ$\\
3 46 25 &  24 42 00 & 20 Sep & 35.89 & $ IZ$\\
3 46 30 &  24 25 00 & 20 Sep & 19.25 & $ IZ$\\
3 46 30 &  24 37 00 & 21 Sep & 30.77 & $ IZ$\\
3 47 20 &  23 25 00 & 21 Sep & 42.25 & $ IZ$\\
3 47 22 &  22 39 40 & 13 Feb & 87.48 & $RIZ$\\
3 47 30 &  24 26 00 & 20 Sep & 20.19 & $ IZ$\\
3 47 52 &  24 10 00 & 21 Sep & 12.23 & $ IZ$\\
3 48 00 &  24 00 00 & 21 Sep & 15.39 & $ IZ$\\
3 48 00 &  24 34 00 &  9 Feb & 30.25 & $RIZ$\\
3 48 05 &  23 39 32 &  9 Feb & 31.24 & $RIZ$\\
3 48 05 &  24 45 00 & 20 Sep & 40.76 & $ IZ$\\
3 48 15 &  23 32 30 & 21 Sep & 38.55 & $ IZ$\\
3 48 15 &  24 28 00 & 21 Sep & 27.06 & $ IZ$\\
3 48 19 &  25 32 00 & 11 Feb & 86.85 & $RI$ \hspace*{0.17cm}\\
3 48 20 &  23 45 00 & 20 Sep & 28.62 & $ IZ$\\
3 48 30 &  23 57 00 & 21 Sep & 22.87 & $ IZ$\\
3 48 40 &  22 42 00 & 13 Feb & 88.07 & $RIZ$\\
3 48 40 &  24 45 00 & 20 Sep, 13 Feb & 44.27 & $ IZ$\\
3 48 45 &  23 40 00 & 20--21 Sep & 36.15 & $ IZ$\\
3 48 57 &  24 18 00 & 21 Sep & 28.84 & $ IZ$\\
3 49 15 &  24 31 00 & 20 Sep & 38.97 & $ IZ$\\
3 49 20 &  23 32 00 & 21 Sep & 47.48 & $ IZ$\\
3 49 20 &  24 45 00 & 20 Sep & 49.54 & $ IZ$\\
3 49 35 &  23 55 00 & 21 Sep & 37.40 & $ IZ$\\
3 49 41 &  25 15 40 & 11 Feb & 77.72 & $RI$ \hspace*{0.17cm}\\
3 50 00 &  24 28 00 & 20 Sep & 46.03 & $ IZ$\\
3 50 15 &  23 40 00 & 21 Sep & 52.18 & $ IZ$\\
3 50 25 &  24 05 00 & 21 Sep & 46.83 & $ IZ$\\
3 50 30 &  23 50 00 & 21 Sep & 50.94 & $ IZ$\\
3 53 30 &  23 40 00 & 11 Feb & 93.29 & $RI$ \hspace*{0.17cm}\\
        &           &      &       &      \\
\hline
\end{tabular}
\begin{itemize}
 \item[$^{\rm{a}}$] Separation from the cluster center (3$^{\rm{h}}$ 
47$^{\rm{m}}$, +24\degr 7\arcmin).
\end{itemize}
\end{center}
\end{table}

\begin{figure}
\resizebox{8.8cm}{!}{\includegraphics{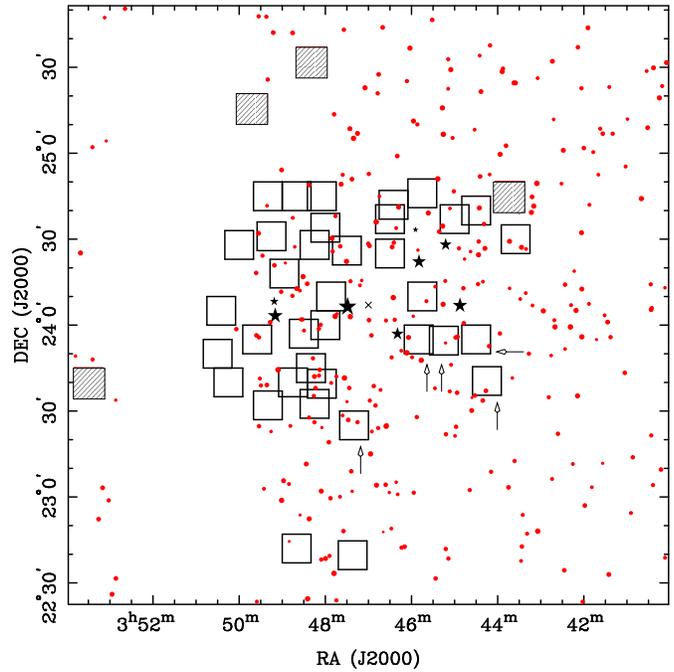}}
\caption[]{\label{campos} Location of our fields (squares) within 
$3^{\circ}.5 \times 3^{\circ}.5$ of the Pleiades area. Open squares 
stand for those fields observed with $IZ$ filters while the shaded 
squares depict the four fields observed with $RI$ filters. Central 
coordinates (indicated with a cross) are 3$^{\rm{h}}$ 47$^{\rm{m}}$, 
+24$^{\circ}$ 7$'$ (Eq. 2000). The five fields which may have some 
amount of extra reddening according to the CO contours shown by Breger 
(\cite{breger87}) are indicated with an arrow. Filled star symbols 
stand for stars brighter than 6$^{\rm th}$\,magnitude, and filled 
circles for proper motion M members (Hambly et al. 1993) with $I$ 
magnitudes in the range 13--18. The vertical gap in the M star 
distribution around 3$^{\rm{h}}$ 51$^{\rm{m}}$ is due to the fact 
that there were no overlaps between the first and second epoch plates 
used by the authors, causing the lack of proper motion measurements 
for stars in that strip. The relative brightness is represented by 
symbol diameters. North is up and East is left}
\end{figure}

All of our CCD survey was carried out during 1996 in two campaigns 
which took place on February 9--12 and on September 19--20. We used 
the TEK (1024$\times$1024 pixel$^2$) detector mounted on the prime 
focus of the telescope, with a field of view of 10$\times$10\,arcmin$^2$. 
A total of 40 fields, with center coordinates listed in Table~\ref{pholog}, 
were observed with the Harris $(R)I$ and RGO $Z$ broad-band filters 
providing a total survey area of 1.05\,deg$^2$ ($\sim$\,6.5\% \ of the 
whole cluster). Most of these fields were selected to avoid very bright 
stars and were located within 1\,deg of the innermost region of the 
Pleiades (see Table~\ref{pholog}), where the population of M dwarf 
proper motion members is much larger than in outer areas (Hambly et al. 
\cite{hambly93}; Bouvier et al. \cite{bouvier98}). If as expected less 
massive Pleiades members show a similar spatial distribution within the 
cluster our survey should be able to detect a large number of BD candidates. 
In Fig.~\ref{campos} the location of all the frames obtained during the 
two observing runs is presented to scale. None of them (except for five 
fields indicated in the figure) falls within the small region southwest 
of the cluster center which is well known to suffer from high absorption 
(van Leeuwen \cite{leeuwen83}; Breger \cite{breger87}; Stauffer \& Hartmann 
\cite{stauffer87}). A small fraction (10.5\%) of our covered area was 
imaged in the three bands while a similar area was observed only in $RI$ 
filters. Weather conditions were always photometric, and the seeing 
oscillated between 1\arcsec \ and 1\farcs5. Typical exposure times ranged 
from 10\,min for the $R$ filter to 5\,min for the $I$ and $Z$-bands.

We adopted the $IZ$ broad-band filters for several reasons. One of our 
goals was to detect objects fainter and less massive than the two cluster 
BDs Teide\,1 and Calar\,3 (M8, $I\,\sim\,19$, $R-I\,\sim\,2.6$, 
$\sim\,0.055\,M_{\odot}$, Rebolo et al. \cite{rebolo95}; Mart\'\i n et 
al. \cite{martin96}). Theoretical evolutionary models (which do not 
include grain formation in very cool atmospheres) predict that these 
objects become much redder with colours $(R-I)$~$\ge$~3 (Chabrier et al. 
\cite{chabrier96}). Thus, they might be extremely faint in $R$ wavelengths, 
greatly hindering their detection. On the other hand, field stars do 
exhibit a turn-off in $(R-I)$ at around M7 spectral type, with stars of 
later types having bluer colours (Bessell \cite{bessell91}). The fluxes 
and colours of the Pleiades BDs fainter than Teide\,1 and Calar\,3 are 
unknown, but we expect them to have spectral energy distributions which 
resemble those of the coolest objects in the field. It could turn out 
that the $(R-I)$ colour is no longer useful to discriminate low luminosity 
cluster members from field objects. The $(I-J)$ colour, however, gets 
monotonically redder for lower temperatures (both for observed and 
theoretical predictions), implying that the slope of the spectral 
pseudocontinuum between $I$ and $J$ wavelengths clearly increases. 
As the $Z$ filter is centered at 920\,nm, we expect a similar behaviour 
with $I$ and $Z$. Although the efficiency of the CCD drops considerably 
in the $Z$-band, this effect is compensated by the increased brightness 
of BDs at these near-IR wavelengths. The $(I-Z)$ colour has been shown to 
be a useful discriminant for Pleiades BDs by Cossburn et al. 
(\cite{cossburn97}).

Other photometric searches for substellar objects in the Pleiades carried 
out with $R$ and $I$ (Jameson \& Skillen \cite{jameson89}; Zapatero Osorio 
et al. \cite{osorio97b}, Paper\,I) provide a high number of mid- and 
late-M stars that do not belong to the cluster and are contaminating the 
surveys. It is desiderable to find a strategy which avoids these field 
contaminants and facilitates a more efficient tool for detecting true 
members. In Paper\,I the success rate was only 25\%: two out of the eight 
proposed cool candidates have been confirmed as {\sl genuine} Pleiades 
BDs (Rebolo et al. \cite{rebolo96}). The authors argue that this was due 
to the detection of reddened late-M dwarfs (Zapatero Osorio et al. 
\cite{osorio97c}, Paper\,II). The use of longer wavelength filters would 
help to jump over this obstacle.

\begin{figure}
\resizebox{8.8cm}{!}{\includegraphics{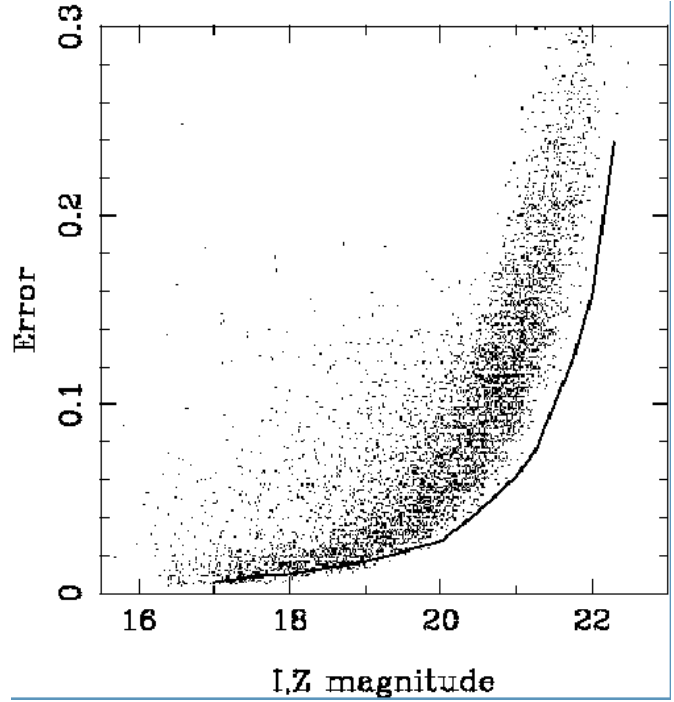}}
\caption[]{\label{error} Photometric errors as a function of observed $I$ 
(full line) and $Z$ (dots) magnitudes. The observed dispersion in the errors 
for the $I$-band is similar to that of the $Z$-band}
\end{figure}

Raw frames were processed using standard techniques within the 
IRAF\footnote{IRAF is distributed by National Optical Astronomy 
Observatories, whcih is operated by the Association of Universities 
for Research in Astronomy, Inc., under contract with the National 
Science Foundation.} (Image Reduction and Analysis Facility) environment, 
which included bias subtraction, flat-fielding and correction for bad pixels 
by interpolation with values from the nearest-neighbour pixels. The 
photometric PSF fitting analysis was carried out using routines within 
DAOPHOT, which provides image profile information needed to discriminate 
between stars and galaxies. Instrumental $RI$ magnitudes were corrected 
for atmospheric extinction and transformed into the $RI$ Cousins system 
using observations of standard stars from Landolt's (\cite{landolt92}) 
list. Special care was taken in including red standard stars in order 
to ensure a reliable transformation for the reddest candidates: the 
field SA\,98 contains many photometric standards covering colours from 
A0 to M7 spectral type. The calibration of $Z$ magnitudes required more 
observational effort as there are no real data for standards available 
in the literature. We have not performed an absolute flux calibration 
for this filter, but obtained $(I-Z)$ colours with respect to a given 
spectral type. Using the same Landolt fields as observed through the 
other two filters at culmination (airmass~=~1.1), we set $Z$~=~$I$ for 
those standard stars with $(R-I)$~$\sim$~0 (A0-type). The adopted $(I-Z)$ 
colours are shown in Table~\ref{sa98}. Observations of these fields 
at different elevations allowed us to correct $Z$ instrumental magnitudes 
for atmospheric extinction. Errors for $Z$ instrumental magnitudes as 
provided by IRAF routines are plotted in Fig.~\ref{error}. The best power 
law fit to the errors in $I$ for the bulk of data is superimposed in the 
figure for comparison. Summarizing, uncertainties in the INT photometry 
range from $\le$0.05\,mag at $I$,$Z$~$\sim\,20.5$, 19.7 to about 0.15\,mag 
at 22, 21\,mag, respectively.

\begin{table}
\caption[]{\label{sa98} Adopted $(I-Z)$ colours for the photometric 
standard stars used in the calibrations}
\begin{center}
\begin{tabular}{lccc}
\hline
        &                  &                  &                  \\
\multicolumn{1}{l}{{\bf SA Star}} &
\multicolumn{1}{c}{{\bf $(R-I)^{a}$}} &
\multicolumn{1}{c}{{\bf $I^{a}$}}  &
\multicolumn{1}{c}{{\bf $(I-Z)^{b}$}} \\
\hline
        &                  &                  &                  \\
98\,642 & 0.393$\pm$0.002  & 14.594$\pm$0.026 &   0.15$\pm$0.09  \\
98\,650 & 0.086$\pm$0.002  & 12.105$\pm$0.003 &   0.02$\pm$0.02  \\
98\,652 & 0.339$\pm$0.024  & 14.201$\pm$0.025 &   0.07$\pm$0.04  \\
98\,653 & 0.008$\pm$0.001  &\ \,9.522$\pm$0.002 &   0.00$\pm$0.02  \\
98\,670 & 0.653$\pm$0.001  & 10.555$\pm$0.003 &   0.23$\pm$0.03  \\
98\,671 & 0.494$\pm$0.004  & 12.315$\pm$0.006 &   0.13$\pm$0.03  \\
98\,675 & 1.002$\pm$0.002  & 11.314$\pm$0.004 &   0.39$\pm$0.03  \\
98\,676 & 0.673$\pm$0.022  & 11.716$\pm$0.005 &   0.24$\pm$0.02  \\
98\,682 & 0.352$\pm$0.003  & 13.032$\pm$0.006 &   0.26$\pm$0.05  \\
98\,685 & 0.280$\pm$0.002  & 11.384$\pm$0.005 &   0.05$\pm$0.02  \\
98\,L5  &  2.60$\pm$0.04   &  12.05$\pm$0.20  &    1.1$\pm$0.2  \\
\hline
\end{tabular}
\begin{itemize}
 \item[$^{a}$] $RI$ magnitudes and their errors taken from Landolt 
(\cite{landolt92}).
 \item[$^{b}$] 1-$\sigma$ errors come from uncertainties in $I$-band 
and the dispersion in the calibration.
\end{itemize}
\end{center}
\end{table}

\begin{figure}
\resizebox{8.8cm}{!}{\includegraphics{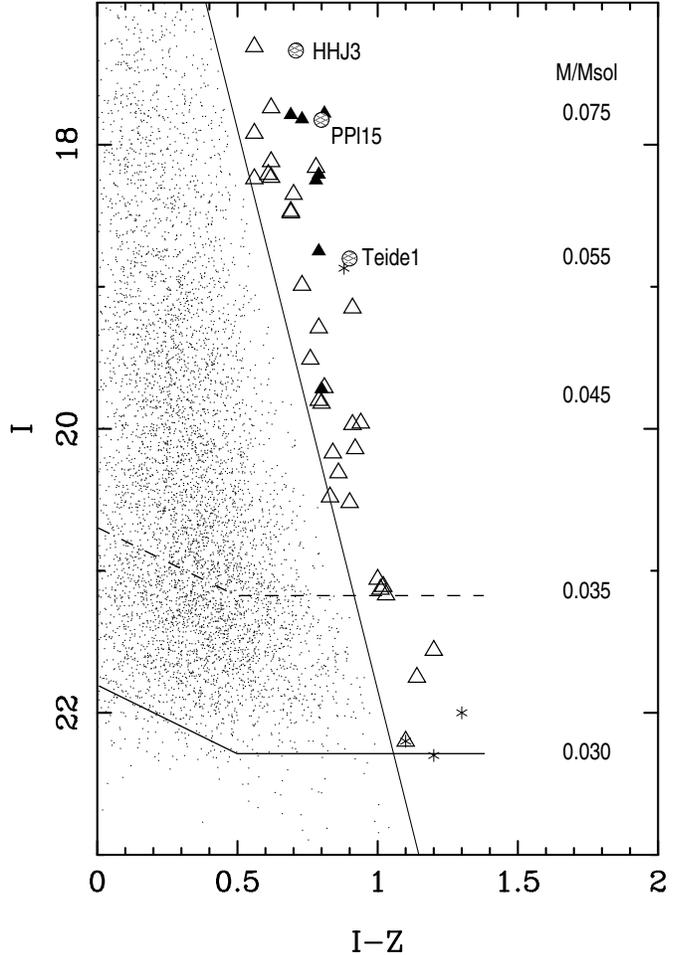}}
\caption[]{\label{iz} $IZ$ colour-magnitude diagram for our 1.05\,deg$^2$ 
survey in the Pleiades. $Z$ magnitudes are not on a standard system (see 
text for details). Previous known members are labelled along with the 
completeness (dashed line) and limiting (full horizontal line) magnitudes. 
Suspected extended objects are shown with asterisks, and the seven 
candidates previously studied in Zapatero Osorio et al. (1997a) are 
indicated with filled triangles. Masses according to the NG Chabrier 
et al.'s (1996) model for solar metallicity and 120\,Myr are labelled 
on the right side}
\end{figure}

We present in Fig.~\ref{iz} the resulting $I$ vs. $(I-Z)$ diagram where 
data for the Pleiads HHJ\,3, PPl\,15 and Teide\,1 (which are present in 
three of our fields) are combined with the new observations. We remark 
that $Z$ magnitudes are not on a standard system. Completeness 
and limiting magnitudes of our survey were derived following the same 
procedure described in Stauffer et al. (\cite{stauffer94}) and Paper\,I. 
We estimate them to be $I$,$Z \sim\,21$, 20.5 for completeness and 22.2, 
21.5 for the limit. These values are indicated in the figure. Because 
there are almost no measurements in the $Z$-band of other cluster members, 
it is rather difficult to establish the separation between Pleiads and 
field objects in our diagram. However, we have made an attempt to separate 
these two kinds of objects by plotting a straight line in Fig.~\ref{iz} 
which is parallel to the photometric sequence defined by HHJ\,3, PPl\,15 
and Teide\,1 and shifted 0.3\,mag towards the blue. For bluer colours 
than those indicated by the line, the number of detections increases very 
drastically, probably indicating that these are field objects. On the 
other hand, the photometric dispersion observed in other optical and 
infrared colours among low-mass proper motion members is about 
0.6--0.7\,mag (Steele \& Jameson \cite{steele95}; Mart\'\i n et al. 
\cite{martin96}). Given the proximity in wavelength of the $I$ and $Z$ 
filters it is expected that this dispersion becomes smaller and therefore, 
the adopted shift takes into account possible binarity effects. For example, 
PPl\,15 was first claimed to be a photometric binary in Paper\,II and 
actually it has been confirmed as a double-lined spectroscopic binary 
with nearly identical components (Basri \& Mart\'\i n \cite{basri98}). 
Those objects fainter than HHJ\,3 and PPl\,15 and located on the right 
side of the straight line are considered our best BD candidates. There are 
43 BD candidate members of the Pleiades in total, plus one (slightly 
brighter) very low-mass candidate stellar member of the cluster. A better 
definition of the true location of this line should be derived after IR 
photometry and spectroscopy are obtained for the candidates (Zapatero 
Osorio et al. \cite{osorio98b}).

\begin{table*}
\caption[]{\label{data} Coordinates and photometry for the candidates}
\begin{center}
\begin{tabular}{llcclcccl}
\hline
                   &          &           &          &          &       
&      &      &  \\
\multicolumn{1}{c}{IAU Name$^{\rm{a}}$} &
\multicolumn{1}{c}{Abridged} &
\multicolumn{2}{c}{RA (J2000) DEC} &
\multicolumn{1}{c}{Epoch} &
\multicolumn{1}{c}{$I$} &
\multicolumn{1}{c}{$I-Z$} &
\multicolumn{1}{c}{$R-I$} &
\multicolumn{1}{c}{Other names$^{\rm{b}}$} \\
\multicolumn{1}{c}{} &
\multicolumn{1}{c}{name} &
\multicolumn{1}{c}{($^{\rm{h \ \ m \ \ s}}$)} &
\multicolumn{1}{c}{(\degr \ \ \arcmin \ \ \arcsec)} &
\multicolumn{1}{c}{} &
\multicolumn{1}{c}{} &
\multicolumn{1}{c}{} &
\multicolumn{1}{c}{} &
\multicolumn{1}{c}{} \\
\hline
                   &          &           &          &          &       
&      &      &  \\
RPL J034741+2244.5 & Roque 48 & 3 47 41.3 & 22 44 33 & 1996.123 & 17.31 
& 0.56 & 1.73 &  \\
                   & HHJ 3$^{\rm{c}}$ & 3 48 50.4 & 22 44 30 & 1989.9 
& 17.33 & 0.71 & 2.21 & \\
RPL J034904+2333.7 & Roque 47 & 3 49 04.8 & 23 33 40 & 1996.726 & 17.74 
& 0.62 &      &  \\
RPL J034723+2242.6 & Roque 17 & 3 47 23.9 & 22 42 38 & 1996.123 & 17.78 
& 0.81 & 2.31 &  \\
RPL J034739+2436.4 & Roque 16 & 3 47 39.0 & 24 36 22 & 1996.112 & 17.79 
& 0.69 & 2.24 & CFHT-PL-11\\
RPL J034541+2354.2 & Roque 15 & 3 45 41.2 & 23 54 10 & 1996.723 & 17.82 
& 0.73 &      & PPl\,1 \\
                   & PPl 15$^{\rm{c}}$ & 3 48 04.8 & 23 39 32 &          
& 17.82 & 0.80 & 2.28 &  \\
RPL J034953+2359.0 & Roque 46 & 3 49 53.7 & 23 59 01 & 1996.726 & 17.92 
& 0.56 &      &  \\
RPL J034738+2238.7 & Roque 44 & 3 47 38.7 & 22 38 41 & 1996.123 & 18.12 
& 0.62 & 2.01 &  \\
RPL J034813+2428.1 & Roque 43 & 3 48 13.8 & 24 28 03 & 1996.726 & 18.16 
& 0.78 &      &JS\,1, PPl\,3\\
RPL J034643+2424.9 & Roque 14 & 3 46 43.0 & 24 24 51 & 1996.723 & 18.21 
& 0.79 &      &  \\
RPL J034939+2334.9 & Roque 42 & 3 49 39.3 & 23 34 55 & 1996.726 & 18.21 
& 0.61 &      &  \\
RPL J034644+2435.0 & Roque 41 & 3 46 44.4 & 24 35 00 & 1996.726 & 18.23 
& 0.62 &      &  \\
RPL J035000+2428.3 & Roque 40 & 3 50 00.3 & 24 28 16 & 1996.723 & 18.24 
& 0.56 &      &  \\
RPL J034550+2409.1 & Roque 13 & 3 45 50.6 & 24 09 03 & 1996.723 & 18.25 
& 0.78 & 2.32 &  \\
RPL J034637+2435.0 & Roque 38 & 3 46 37.7 & 24 35 02 & 1996.726 & 18.35 
& 0.70 &      &  \\
RPL J034819+2425.2 & Roque 12 & 3 48 19.0 & 24 25 12 & 1996.726 & 18.47 
& 0.69 &      & NPL\,36 \\
RPL J034844+2421.3 & Roque 37 & 3 48 44.1 & 24 21 18 & 1996.726 & 18.48 
& 0.69 &      &  \\
RPL J034712+2428.5 & Roque 11 & 3 47 12.0 & 24 28 31 & 1996.723 & 18.75 
& 0.79 & 2.68 & NPL\,37 \\
TPL J034718+2422.5 & Teide 1$^{\rm{c}}$  & 3 47 18.0 & 24 22 31 & 1994.85  
& 18.80 & 0.90 & 2.74 &  \\
RPL J034802+2400.1 & Roque 10$^{\rm{d}}$ & 3 48 02.1 & 24 00 03 & 1996.723 
& 18.87 & 0.88 &      &  \\
RPL J034623+2420.6 & Roque 9  & 3 46 23.2 & 24 20 37 & 1996.723 & 18.99 
& 0.73 &      &  \\
RPL J034921+2334.0 & Roque 8  & 3 49 21.1 & 23 34 02 & 1996.726 & 19.15 
& 0.91 &      &  \\
RPL J034340+2430.2 & Roque 7  & 3 43 40.3 & 24 30 11 & 1996.726 & 19.29 
& 0.79 &      & CFHT-PL-24\\
RPL J034957+2341.8 & Roque 6  & 3 49 57.8 & 23 41 50 & 1996.726 & 19.51 
& 0.76 &      &  \\
RPL J034422+2339.0 & Roque 5  & 3 44 22.4 & 23 39 01 & 1996.726 & 19.71 
& 0.81 &      &  \\
RPL J034353+2431.2 & Roque 4  & 3 43 53.5 & 24 31 11 & 1996.726 & 19.72 
& 0.80 & 2.30 &  \\
RPL J034410+2340.3 & Roque 3  & 3 44 10.9 & 23 40 15 & 1996.726 & 19.80 
& 0.79 &      &  \\
RPL J034420+2439.0 & Roque 36 & 3 44 20.8 & 24 39 02 & 1996.723 & 19.82 
& 0.80 &      &  \\
RPL J034737+2429.0 & Roque 34 & 3 47 37.5 & 24 28 59 & 1996.723 & 19.96 
& 0.94 &      &  \\
RPL J034849+2420.5 & Roque 33 & 3 48 49.0 & 24 20 25 & 1996.726 & 19.97 
& 0.91 &      & NPL\,40 \\
RPL J034705+2324.9 & Roque 32 & 3 47 05.8 & 23 24 52 & 1996.726 & 20.14 
& 0.92 &      &  \\
RPL J035032+2408.9 & Roque 31 & 3 50 32.5 & 24 08 53 & 1996.726 & 20.17 
& 0.84 &      &  \\
RPL J035016+2408.5 & Roque 30 & 3 50 16.0 & 24 08 35 & 1996.726 & 20.31 
& 0.86 &      &  \\
RPL J035005+2342.2 & Roque 29 & 3 50 05.9 & 23 42 14 & 1996.726 & 20.48 
& 0.83 &      &  \\
RPL J035020+2408.7 & Roque 28 & 3 50 20.7 & 24 08 41 & 1996.726 & 20.52 
& 0.90 &      &  \\
RPL J034746+2403.7 & Roque 27 & 3 47 46.9 & 24 03 42 & 1996.726 & 21.06 
& 1.00 &      &  \\
RPL J034412+2343.3 & Roque 18 & 3 44 12.6 & 23 43 17 & 1996.726 & 21.11 
& 1.02 &      &  \\
RPL J034849+2245.9 & Roque 26 & 3 48 49.3 & 22 45 51 & 1996.123 & 21.13 
& 1.01 & 1.98 &  \\
RPL J034830+2244.9 & Roque 25 & 3 48 30.6 & 22 44 50 & 1996.123 & 21.17 
& 1.03 & 2.75 &  \\
RPL J034951+2515.5 & Roque 2  & 3 49 51.4 & 25 15 31 & 1996.117 & 21.31 
&      &$\ge$2.1&\\
RPL J034321+2434.7 & Roque 24 & 3 43 21.4 & 24 34 42 & 1996.726 & 21.56 
& 1.20 &      &  \\
RPL J034751+2355.8 & Roque 23 & 3 47 51.0 & 23 55 48 & 1996.726 & 21.75 
& 1.14 &      &  \\
RPL J034321+2432.0 & Roque 22$^{\rm{d}}$ & 3 43 21.3 & 24 32 02 & 1996.726 
& 22.0: & 1.3: &      &  \\
RPL J034327+2433.7 & Roque 21$^{\rm{d}}$ & 3 43 27.8 & 24 33 40 & 1996.726 
& 22.2:& 1.1:&      &  \\
RPL J034843+2240.8 & Roque 20 & 3 48 43.7 & 22 40 46 & 1996.123 & 22.2:
& 1.1:&      &  \\
RPL J034855+2420.2 & Roque 19$^{\rm{d}}$ & 3 48 55.3 & 24 20 09 & 1996.726 
& 22.3:& 1.2:&      &  \\
\hline
\end{tabular}
\begin{itemize}
 \item[$^{\rm{a}}$] ``RPL'' stands for Roque Pleiades, and ``TPL'' for 
Teide Pleiades.
 \item[$^{\rm{b}}$] References: PPl objects from Stauffer et al. 
(\cite{stauffer89}); JS objects from Jameson \& Skillen (\cite{jameson89}); 
NPL objects from Festin (\cite{festin98}); CFHT-PL objects from Bouvier 
et al. (\cite{bouvier98}).
 \item[$^{\rm{c}}$] Coordinates for HHJ\,3 taken from Hambly et al. 
(\cite{hambly93}); for PPl\,15 from Stauffer et al. (\cite{stauffer94}); 
coordinates and $RI$ photometry for Teide\,1 taken from Paper\,I.
 \item[$^{\rm{d}}$] These objects appear slightly extended in the $IZ$ 
images.
 \item[:]           For error bars in the photometry see text. Those 
measurements labelled with a ``:'' have rather large uncertainties.
\end{itemize}
\end{center}
\end{table*}

In Table~\ref{data} we list the names, magnitudes, colours and positions 
for the proposed Pleiades BD candidates. They are named after the 
{\em Roque} Observatory followed by the word {\em Pleiades} and numbered 
according to their decreasing $I$-band apparent magnitude (second column 
of Table~\ref{data}). Hereafter, we will use an abridged version of the 
names which omits the term `Pleiades'. The names of the candidates adopting 
the IAU rules are also provided (first column), where the acronymn ``RPL'' 
stands for {\em Roque Pleiades}. Three of the four faintest candidates 
have slightly larger fwhm than the average value for our frames. 
Presumably this is an indication that they are not a point source. It 
is expected that distant galaxies fainter than $I$~=~21 will begin to 
contaminate the number counts of objects. Those candidates labelled as 
extended are shown with a different symbol in Fig.~\ref{iz}. By reference 
to the reddening map provided in Breger (\cite{breger87}), Roque\,3, 5, 
15, 18 and~32 could suffer from a somewhat enhanced extinction as they 
lay within or very near to the CO contours given by the author.

In addition to the INT data, we have obtained $R$-band photometry for 
five of the candidates at the 2.5\,m Nordic Optical Telescope (BroCam1, 
NOT) on 1996 October 10--11 (Roque\,17, 11 and 4), and at the 1\,m Jacobus 
Kaptein Telescope (JKT) on 1996 September 12--13 (Roque\,16 and 13), both 
telescopes at the ORM. The CCDs used were a Tektronix 1024$\times$1024 
providing fields of view of 3.0 and 5.5\,arcmin$^2$, respectively. 
Exposure times were typically 15\,min at NOT and 30\,min at JKT. Landolt's 
(\cite{landolt92}) standard stars were observed just before and after the 
targets. Reduction of the raw frames and photometry of the candidates has 
been performed as described above. Uncertainties in $R$ magnitudes range 
from 0.07\,mag for the brightest objects to 0.15\,mag for the faintest 
ones. Considering the $R-I$ photometry from Table~\ref{data} and from 
other deep surveys (Paper\,I; Bouvier et al. \cite{bouvier98}) Roque\,44 
and Roque\,26 are not likely to be Pleiades members as they seem to deviate 
towards bluer colours from the sequence defined by other candidates.

\begin{figure}
\resizebox{8.8cm}{!}{\includegraphics{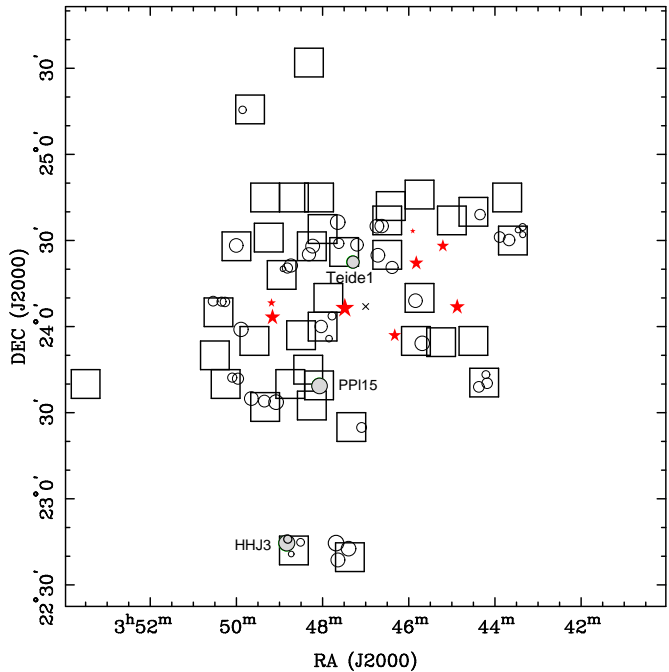}}
\caption[]{\label{loccand} Location of our candidates within the fields 
observed in our survey covering 1.05\,deg$^2$. As in Fig.~\ref{campos} 
central coordinates (3$^{\rm{h}}$ 47$^{\rm{m}}$, +24$^{\circ}$ 7$'$, 
Eq. 2000) are indicated with a cross. Filled star symbols outline stars 
brighter than 6$^{\rm th}$\,magnitude. The relative brightness is 
represented by symbol diameters. North is up and East is left}
\end{figure}

Astrometry for all Roque objects has been performed by the triangles 
fitting method using the APM Sky Catalogue. Several stars close to every 
candidate were identified and they served as a reference for the astrometric 
calibration. Coordinates are accurate to approximately $\pm$2\arcsec. 
The location of our candidates in the surveyed area is depicted in 
Fig.~\ref{loccand}. Their distribution around the cluster center appears 
quite homogeneous. However, we note that the number of fields (9) with 
$\geq$\,3 BD candidates is surprisingly large compared to the expectations 
from a random distribution. The study of possible spatial inhomogeneities 
within the cluster still awaits membership confirmation. Seven of the 
Roque BD candidates have also been identified in other surveys. The last 
column of Table~\ref{data} gives cross-identifications. Our $I$ magnitudes 
seem to be on average 0.25\,mag brighter than those available in the most 
recent literature. This is likely due to an effect of the colour-dependence 
of the Harris filters we used in our observations; although a red standard 
star was considered, it is poorly calibrated and consequently does not 
provide an accurate determination of the colour-term in the photometric 
calibration. Cossburn et al. (\cite{cossburn98}) have found that the 
colour-term for the transformation from Harris $I$ to Cousins is indeed 
rather significant for very red objects. In the case of 
Roque\,33 (NPL\,40) the difference found is $-0.58$\,mag which might be 
due to contamination from a nearby very bright star (and saturated in our 
frames). Figure~\ref{cartas} provides the $I$-band finder charts 
(2\arcmin $\times$ 2\arcmin \ in extent) for all Roque objects ordered 
as listed in Table~\ref{data}.

According to the ``NextGen'' (NG) theoretical evolutionary models of 
Chabrier et al. (\cite{chabrier96}), and adopting solar metallicity, 
an age of 120\,Myr (Basri et al. \cite{basri96}; Mart\'\i n et al. 
{\cite{martin98}; Stauffer et al. \cite{stauffer98}) and a distance of 
127\,pc for the Pleiades cluster, our survey has detected objects in 
the mass interval from roughly 0.08\,$M_{\odot}$ down 
to 0.03\,$M_{\odot}$. The completeness magnitudes correspond to 
0.035\,$M_{\odot}$ as indicated in Fig.~\ref{iz}. Chabrier et al.'s models 
provide absolute magnitudes as a function of mass, metallicity and age 
obtained by direct integration of theoretical atmospheres which do not 
incorporate grain formation and dust absorption (Allard et al. 
\cite{allard97}). However, the effects of condensation become important 
for temperatures cooler than about 2500\,K (Tsuji et al. \cite{tsuji96}; 
Jones \& Tsuji \cite{jones97}), a temperature range partially covered 
by our survey. Preliminary computations by Baraffe (private communication) 
show that models considering dust formation and opacities predict brighter 
$I$ magnitudes and subsequently slightly lowers the mass determination 
by $\sim\,8$\,\%.

Membership and therefore the real nature of our candidates on the basis 
of $JHK$ photometry and spectroscopy and the Pleiades mass function will 
be addressed in a forthcoming paper (Zapatero Osorio et al. 
\cite{osorio98b}). Seven of them (Roque\,17, 16, 15, 14, 13, 11 and~4) 
with $I$ magnitudes in the range 17.8--19.5 (masses in the interval 
0.08--0.045\,$M_{\odot}$) have already been studied to some extent by 
Zapatero Osorio et al. (\cite{osorio97a}). They are shown in Fig.~\ref{iz} 
with a different symbol. The authors conclude that given their $K$ 
magnitudes, radial velocities, spectral types and weakness of some 
atomic features these candidates should be considered as Pleiades 
members. The number of remaining candidates in our $IZ$ survey deserve 
further investigation as there are large enough to ensure that follow-up 
observations will confirm more Pleiades substellar objects. Among the 
faintest ones, there could be BDs with masses as low as 0.03\,$M_{\odot}$. 
These studies will make it possible to derive the cluster mass function 
well into the substellar regime.

\section{Conclusions}

We have imaged about 1\,deg$^2$ in the central region of the Pleiades 
young cluster using $IZ$ broad-band filters, and identified more than 40 
brown dwarf candidates with $I$ magnitudes ranging from 17.5 to 22 
(completeness is given by $I \sim$21\,mag). This corresponds to a mass 
interval starting roughly at the substellar mass limit down to 
0.03\,$M_{\odot}$ according to recent non-dusty evolutionary models by 
Chabrier et al. (\cite{chabrier96}). Follow-up IR and spectroscopic 
observations of seven of these candidates (Zapatero Osorio et al. 
\cite{osorio97a}) have revealed that they are very likely to be cluster 
brown dwarfs. Further data for the remaining candidates should definitely 
establish or refute cluster membership in the Pleiades. These studies 
will provide a solid foundation for deriving the cluster substellar mass 
function.

\begin{acknowledgements}
We thank the referee, J. Stauffer, for his valuable comments. We also 
thank the Comit\'e Cient\'\i fico Internacional for awarding 25\% \ of 
the International Time at the Observatories of the Canary Islands to 
this project in 1996--97. Partial financial support was provided by 
the Spanish DGES project no. PB95-1132-C02-01. A. M. acknowledges a 
NATO-CNR senior fellowship. 
\end{acknowledgements}

\begin{figure*}
\caption[]{\label{cartas} Finder charts (2\arcmin $\times$ 2\arcmin \ in 
extent) taken from the $I$-band images for all Roque candidates. North is 
up and East is left}
\end{figure*}




\end{document}